# Mobility Inequity and Risk Response After Hurricane Helene: Evidence from Real-Time Travel and Social Sentiment Data

Qian He*, Zihui Ma, Songhua Hu, Behnam Tahmasbi

**Authors:**

Qian He, PhD, AICP: Assistant Professor, Department of Geography, Planning, and Sustainability, Rowan University, NJ, USA: Email: he@rowan.edu

Zihui Ma, PhD: Faculty Fellow, Center for Urban Science and Progress, New York University, Brooklyn, NY 11201, Postdoctoral Research Associate, Department of Civil and Environmental Engineering, University of Maryland, College Park, MD 20742, USA Email: zihuima@nyu.edu

Songhua Hu, PhD: Assistant Professor, Department of Architecture and Civil Engineering, City University of Hong Kong, Kowloon, Hong Kong SAR, Email: songhuhu@cityu.edu.hk

Behnam Tahmasbi: Department of Civil and Environmental Engineering, University of Maryland, College Park, MD 20742, USA: Email: behnamt@umd.edu

*Corresponding Author\**

Mailing Address: 201 Mullica Hill Rd, Glassboro, NJ, US, 08028

**Acknowledgement:**

The authors are grateful for the help from Mr. Greg Tune at the American Red Cross in accessing Dwelling Unit Damage Assessment data. This work is partially supported by the Faculty Startup Fund (Q.H.) from the School of Earth and Environment, College of Science and Mathematics, at Rowan University.

**Abstract:**

Hurricanes severely disrupt infrastructure and restrict access to essential services. While the physical impacts on post-disaster mobility are well studied, less is known about how individual travel behaviors change during and after disasters, and how these responses are shaped by social and geographic disparities. This study examines mobility patterns following Hurricane Helene, a Category 4 storm that struck southeastern U.S. states on September 26, 2024, causing over 230 fatalities. Using anonymized GPS mobility data, hurricane severity metrics, and county-level social media sentiment, we examine shifts in travel behavior and their implications for equity. We ask two questions: How do post-hurricane mobility patterns reflect community vulnerability and adaptive capacity? and How do sociodemographic conditions and public sentiment factors shape the direction and extent of mobility change? Results from robust linear and ordered logistic regressions indicate that evacuation orders increase mobility; however, severe storm conditions, particularly high wind speeds, can limit travel. Communities with lower incomes, located in rural areas, and with higher percentages of Black populations exhibit the steepest declines in mobility, suggesting resource constraints and infrastructural barriers, while wealthier, urban, and higher-education areas maintain greater flexibility. Results also show that positive social sentiment is associated with higher mobility and a greater likelihood of increased travel during the hurricane. Our findings highlight the need to address structural barriers and social conditions in post-disaster mobility and disaster response.

## Introduction

Climate-induced disasters, particularly hurricanes, disrupt community infrastructure and restrict access to essential services. Flooding, traffic congestion, and damaged transportation networks can severely limit residents' ability to reach destinations such as healthcare facilities, grocery stores, and emergency shelters (T. Logan et al., 2019; Swanson & Guikema, 2024; Washington et al., 2024). Research has shown that evacuation and recovery decisions are influenced by risk perception, socioeconomic resources, and the condition of infrastructure. Rural areas often experience partial or staggered evacuations due to limited road capacity and long travel distances (Sadri et al., 2018), while households with health or mobility constraints are less likely to evacuate unless supported by tailored services (Bethel et al., 2011; Hostetter & Naser, 2022; Van Willigen et al., 2002). Other studies link evacuation likelihood to anticipated disruptions in critical services, such as electricity or medical care (Lamadrid et al., 2025), and emphasize the roles of income, access to medication, and transportation in shaping evacuation timing (Diaz et al., 2023). Mobile phone data has recently expanded this literature by capturing real-time evacuation flows and bottlenecks, complementing survey-based findings (X. Li et al., 2024).

Despite these advances, important gaps remain. Much of the existing work focuses on whether households evacuate but less on how mobility behavior evolves during and after disasters, and how these shifts are shaped by social and geographic disparities. Further, while mobility data reveal when and where populations move, they provide limited insight into whether trips represent adaptive recovery or distress-driven displacement. Studies that integrate behavioral data with public sentiment are rare. With Hurricane Helene, a Category 4 storm that struck on September 26, 2024, devastating five southeastern states and causing over 230 fatalities, there is an urgent opportunity to deepen understanding of these dynamics. Early evidence suggests that inland communities with less hurricane experience exhibited more constrained adaptive mobility than coastal areas, reflecting differences in infrastructure, socioeconomic disadvantage, and risk perception (Yao et al., 2025).

This study analyzes post-hurricane mobility using two complementary data sources: SafeGraph's anonymized GPS records and sentiment analysis of geolocated posts on platform X (formerly Twitter). To account for disaster intensity, we integrate data from the National Weather Service on precipitation and maximum wind speeds. We ask: (1) How do post-hurricane mobility patterns reflect community vulnerability and adaptive capacity? (2) How do sociodemographic conditions and public sentiment factors shape the direction and extent of mobility change? Our results show that evacuation orders significantly increased mobility, underscoring the importance of timely directives. At the same time, storm severity reduced the likelihood of increased trips, revealing the constraining effects of hazardous conditions. Demographic disparities were evident: Black and rural communities displayed lower mobility responses, while higher-income and more educated populations showed greater adaptive adjustments. Sentiment patterns mirrored these findings, where we observed that distress was associated with reductions in mobility, while positive sentiment was linked to higher mobility and a greater likelihood of increased travel. Our findings show that mobility shifts cannot be read as simple markers of resilience or vulnerability. By integrating sentiment data, we distinguish movements driven by distress from those reflecting adaptive capacity. This approach reveals layered patterns of vulnerability across communities, advances understanding of post-disaster resilience, and helps identify populations most at risk of mobility constraints to inform more equitable disaster response.

**Background**

The impacts of disasters on mobility and travel behavior have received substantial attention, particularly in the context of hurricanes. Hurricanes disrupt transportation infrastructure, significantly affecting residents' ability to access essential services, including healthcare, groceries, and emergency shelters (Q. Wang & Taylor, 2014). The decision-making processes underlying evacuation during hurricanes involve complex interactions among perceived risks, socioeconomic status, infrastructure conditions, and the effectiveness of public communication (Lindell et al., 2005). Notably, evacuation behaviors exhibit substantial heterogeneity, influenced by individual and community-level variables, including prior experiences, trust in authorities, and the availability of evacuation routes and shelters (Dash & Gladwin, 2007). Post-disaster mobility and recovery further underscore existing socioeconomic and racial disparities (Deng et al., 2021). People from historically marginalized social statuses and lower-income populations often encounter more significant barriers to evacuation, access to resources, and post-disaster recovery (Best et al., 2023; Ferreira et al., 2024; Fothergill & Peek, 2004; Logan et al., 2016). These disparities are exacerbated by spatial inequalities, as communities in rural or less densely populated areas typically face greater challenges due to limited transportation infrastructure, heightened car dependency, and fewer available services (Adey, 2016). Fothergill and Peek (2004) demonstrated how people of constrained socioeconomic conditions and limited resources in the aftermath of disasters disproportionately impacts already marginalized population groups, perpetuating cycles of vulnerability and reduced adaptive capacity. Fussell et al. (2010) highlight a slow return migration among Black residents compared to white residents in the 14 months after Hurricane Katrina. That disparity primarily stemmed from greater housing damage, which is not studied in the context of transportation access or evacuation capacity (Fussell et al., 2010). Recent work shows that residential location alone underestimates flood exposure, as mobility patterns extend risk into flood-prone areas (B. Li et al., 2024). Mobility-based exposure measures reveal that Black, Asian, lower-income, and less-educated populations are disproportionately exposed due to their daily activities, which highlight the disparities not captured by residence-based assessments.

However, other research indicates that evacuation plans often assume private vehicle access, leaving transit-dependent, low-income residents—often people of color—stranded when public transportation fails or is unavailable (Litman, 2006; Lui et al., 2006; Naghawi & Wolshon, 2010). Furthermore, research by Sadri et al. stresses that socioeconomic factors, including income and race, strongly predict post-disaster mobility patterns, with economically disadvantaged and minority communities often exhibiting restricted travel and slower recovery (Sadri et al., 2018). Additional studies have explored the psychological and behavioral dimensions of disaster-induced mobility. For instance, research finds that higher education levels can paradoxically reduce evacuation propensity, as educated individuals may possess a higher awareness of disaster preparedness measures, thereby opting to shelter in place (Hasan et al., 2011). Eisenman et al. (2007) studied evacuees from New Orleans shelters following Hurricane Katrina, with a primary focus on low-income, minority communities. They found that decisions to evacuate were strongly shaped by access to transportation and shelter, clarity and trustworthiness of risk communication, and the role of social networks (Eisenman et al., 2007). Barriers, including a misperception of danger, a lack of credible information, and economic constraints, prevented many from evacuating despite the risks. Similarly, risk perception and communication clarity are found to significantly influence evacuation compliance, emphasizing the need for tailored public messages to improve evacuation outcomes among diverse communities (Lindell & Perry, 2012).

Mobile phone data have emerged as a powerful tool for analyzing evacuation flows, displacement, and mobility disruptions during hurricanes in near real-time and with unprecedented spatiotemporal granularity, complementing traditional post-disaster surveys (Yabe et al., 2019a). For example, a seminal study of Hurricane Sandy (2012) utilized high-resolution location records to quantify how the storm perturbed daily travel. Human movement distances immediately after landfall followed a truncated power-law distribution, with significantly shorter ranges than usual (Q. Wang & Taylor, 2014). *Yabe et al. (2019)* examined inter-city movement networks in Puerto Rico after Hurricane Maria (2017) using anonymized mobile phone GPS records(Yabe et al., 2019). They found that counties with greater pre-disaster outbound and inbound travel tended to recover their populations more quickly. (Washington et al., 2024) introduce a data-driven algorithm that infers residents' home locations and flags evacuation events by detecting anomalous spikes in departures relative to historical trends. Applying this method to Hurricanes Matthew (2016) and Irma (2017), they identified which neighborhoods experienced unusual exoduses and the timing of those movements, without relying on predefined evacuation zones. These studies demonstrate the capacity of mobile phone data to rapidly and accurately capture population movements during crises, thereby informing disaster response efforts related to shelter demand, transportation planning, and evacuation compliance, and ultimately enhancing situational awareness across all phases of emergency management (Hu et al., 2024; Swanson & Guikema, 2024).

Theoretically, a decline in mobility following a disaster can be interpreted in two contrasting ways: as an indicator of vulnerability or as an indicator of resilience. In the first scenario, when trip reductions are driven by severe infrastructure damage, road closures, or a lack of transportation options, they can signal heightened risk and entrapment. Individuals in these areas may be unable to reach essential destinations such as hospitals, grocery stores, or evacuation centers, exacerbating their vulnerability. In this scenario, reduced mobility reflects a breakdown in accessibility and emergency response capacity rather than an intentional or positive adaptation. Alternatively, a decrease in trips may reflect self-sufficiency and disaster preparedness, where individuals have the necessary resources and support systems in place to shelter in place without needing to travel. In this way, the reduced travel can be an indicator of resilience. This could indicate strong community resilience, robust emergency planning, and effective resource distribution, reducing the need for unnecessary movement. In such cases, reduced mobility represents stability rather than distress.

Similarly, an increase in trips post-hurricane can be interpreted in contrasting ways, reflecting either emergency-driven displacement or effective resource accessibility: a sign of emergency and displacement, versus a signal of accessibility and coping ability. For example, a surge in mobility may indicate mass evacuations, forced relocations, or desperate efforts to seek aid, suggesting that communities were in significant danger. This form of increased movement reflects heightened distress, instability, and possible systemic failures that necessitate large-scale movement for survival. On the other hand, an increase in trips could indicate that individuals were able to move freely to obtain necessary supplies, access healthcare, and engage in recovery activities. In this context, mobility reflects resilience and adaptive capacity, demonstrating that transportation networks remained functional and that residents could respond effectively to post-disaster challenges. Understanding the contexts underlying these movements is therefore crucial for understanding the causes and conditions of the mobility change. More specifically, public sentiment data can reveal whether mobility stems from fear and distress or from confidence and stability. Social media has become an important tool in disaster research for capturing real-time public sentiment and perceived risk at fine spatial scales. User-generated content, particularly on platforms such as X (formerly Twitter), provides valuable insight into emotional states, social concerns, and levels of institutional trust during extreme events (Beigi et al., 2016; Ma et al., 2024; Tanna et al., 2020). Building on this, recent studies have increasingly leveraged social media data to examine the impacts of hurricanes on urban populations, using data-driven approaches to evaluate preparedness, response, and recovery processes (X. Li et al., 2024; Yabe et al., 2019; Yin et al., 2020). For example, Zhang et al. (2019) used semisupervised learning to classify emotions into four categories and discovered that emotional expression on social media was linked to a user's information-sharing behavior. They found that negative information spread more quickly during the pre- and during periods of 2017 Hurricane Irma. These platforms also enable geolocated analysis of public reactions (Yum, 2021), positioning sentiment analysis as a valuable tool for assessing behavioral responses to disasters.

In the aftermath of a hurricane, mobility patterns can serve as a critical indicator of community resilience. However, interpreting these patterns requires a nuanced understanding, as both reductions and increases in trips can signify either vulnerability or adaptive capacity (Nussbaum, 2003), depending on the context. Several studies have demonstrated that collective emotional tone is closely associated with protective behaviors and mobility-related decisions (Haraguchi et al., 2022; Jiang et al., 2019; Y. Wang & Taylor, 2018). For instance, Wang & Taylor (2018) reveal that fluctuations in sentiment exhibited both lead and lag relationships with changes in population mobility, underscoring the dynamic interplay between emotional response and physical movement. Other studies suggest that a more positive affective tone is correlated with increased engagement in recovery efforts, greater compliance with public directives, and a quicker return to normal travel patterns (L. Li et al., 2020a; Lu, 2024; Young et al., 2024). These responses, however, are often shaped by broader social conditions. Communities with higher baseline optimism, dense social networks, or stronger institutional trust may exhibit significantly different behavioral patterns than those experiencing long-standing disinvestment or systemic marginalization (Bonfanti et al., 2024; D'Souza et al., 2023; Gongora-Svartzman & Ramirez-Marquez, 2022; Tan & Guan, 2021; Xu et al., 2024). Moreover, existing research often treats sentiment and mobility as distinct domains, leaving gaps in our understanding of how emotional and behavioral responses interact with each other.

In this study, we use mobile phone data to monitor spatiotemporal changes in movement patterns, along with social media data, to get a better understanding of the comprehensive view of post-disaster mobility behavior. With our proposed dual framework, this study highlights the limitations of treating mobility trends as universal indicators of resilience or vulnerability without considering broader contextual factors. A decrease in mobility does not always signify distress, just as an increase in trips does not inherently reflect recovery or stability. Instead, understanding why mobility patterns shift and how they correlate with factors such as social vulnerability, infrastructure conditions, and public sentiment is essential for accurately assessing community resilience and guiding disaster response efforts. Our approach emphasizes the complex link between mobility and disaster resilience, showing the importance of context when analyzing post-disaster movement patterns.

**Data and Methodology**

The study area of this paper focuses on regions affected by Hurricane Helene. After landfall, Hurricane Helene tracked inland, causing widespread flooding and damage to infrastructure across *Florida, Georgia, South Carolina, North Carolina, and Tennessee*. The storm resulted in substantial disruptions to transportation and essential services (Yao et al., 2025). These five states consist of communities with diverse geographic conditions, including both coastal and inland counties with varied sociodemographic profiles. To understand the disparity in mobility changes and their association with social sentiment data, we integrate multiple datasets to examine hurricane-related risk exposure, mobility responses, and public sentiment during and after Hurricane Helene. To evaluate local storm intensity, we compiled weather station data from the National Oceanic and Atmospheric Administration (NOAA) Integrated Surface Database (ISD). All active weather stations within the affected states were mapped, and Thiessen polygons were generated to delineate each station's area of spatial influence. We calculated the average daily wind speed (m/s) and precipitation (mm) recorded between September 26 and 28, 2025. This period represents the peak of Hurricane Helene's landfall and inland movement, as observed at the census block group level, and utilizes these two measures as indicators of storm severity across communities.

To assess population movement during the hurricane, we used aggregated and anonymized GPS-based mobility data from SafeGraph. This dataset provides daily counts of trips at the Census Block Group (CBG) level, enabling analysis of mobility disruptions across the region. We compared average daily trip counts during the pre-hurricane period (September 1–26, 2024) to those during the hurricane impact window (September 27–28, 2024). Key indicators include (1) the average number of daily trips per CBG before and during the event, and (2) the percentage change in trip volume relative to the pre-hurricane baseline. These measures capture reductions, surges, or stability in mobility that may reflect evacuation behavior, access disruptions, or early recovery travel. Sociodemographic characteristics of each CBG were drawn from the 2024 American Community Survey (ACS) Five-Year Estimates based on factors identified through our literature review.

To control for government-issued evacuation orders as a controlling factor for mobility movement, we retrieved county-level evacuation order data from the Federal Emergency Management Agency (FEMA) to capture the geographic scope and timing of formal evacuation directives. These data provide a policy context for interpreting observed mobility patterns and travel disruptions. In addition, we obtained residential damage data from the American Red Cross Loss Damage Assessment (LDA) database[1] to assess the localized degrees of damage, which reports the number of dwelling units affected, majorly damaged, and destroyed. We then aggregated these records to the census block group level to measure the local physical impacts of the hurricane.

---

[1]The American Red Cross Loss Damage Assessment (LDA) is a critical initial response procedure for evaluating the extent of destruction following a disaster. Red Cross teams collect data to assess the damage to individual homes and use this information to determine the level of assistance needed and where to deploy resources.

To assess public sentiment and healthcare-related concerns during and after the hurricane, we retrieved and analyzed data from X (formerly Twitter) from September 17 to October 7, 2024, using a series of keywords[2] related to *Hurricane Helene*. We constrained the data to posts originating from the impacted states: FL, GA, NC, SC, TN, and VA. The real-time nature of social media allows users to share experiences and opinions, making it a valuable source for understanding public perceptions during disasters. We applied the Valence Aware Dictionary and sEntiment Reasoner (VADER), a widely used lexicon- and rule-based sentiment analysis tool, to classify each tweet as positive, negative, or neutral. Specifically, the average sentiment score follows the approach of Li et al. (2020) and was computed as:

$$Average\ Sentiment\ Score = \frac{No.\ of\ Positive\ tweets\ -\ No.\ of\ Negative\ tweets}{Total\ No.\ of\ tweets}$$

We then aggregated these scores at the city and county level to assess geographic variations in public outlook.

**Findings and Discussion**

*Spatial Heterogeneity in Post-Helene Mobility*

As illustrated in Figure 1, the top two maps show county-level percentage changes in mobility following the hurricane. The top left map, at a coarser scale on the county level, shows that much of the Southeast experienced declines in mobility (represented by red hues), especially in Florida, while a few localized areas show increases (blue). The right map shows these changes at finer spatial resolution, revealing strong within-state variation. Even within regions with overall declines, pockets of counties still saw sharp increases, suggesting localized recovery or evacuation-related movements. The sharp contrasts between neighboring counties underscore how social and infrastructural contexts shape travel behavior. Relatedly, Figure 2 shows the spatial distributions of storm severity. Precipitation (bottom left) was highest in several inland pockets, especially across parts of Georgia and the Carolinas, while coastal areas experienced more moderate rainfall. Average wind speeds (bottom right) were higher along portions of the Atlantic coast and parts of the Carolinas, with generally lower wind intensity inland. These weather patterns help to contextualize the shifts in mobility. We see that areas with the strongest wind impacts often align with the largest declines in mobility, suggesting that damage or perceived danger may have suppressed travel. Meanwhile, precipitation-heavy areas without major wind damage showed more mixed

---

[2] Keywords applied for Tweet query: (("hurricane helene" OR "hurricnae helen" OR hurricanehelene OR hurricnaehelen OR "tropical storm" OR "tropical cyclone" OR hurricane OR landfall) OR (#helene OR #helen OR #hurricanehelene OR #hurricanehelen))

mobility patterns, indicating that rainfall alone did not uniformly restrict movement.

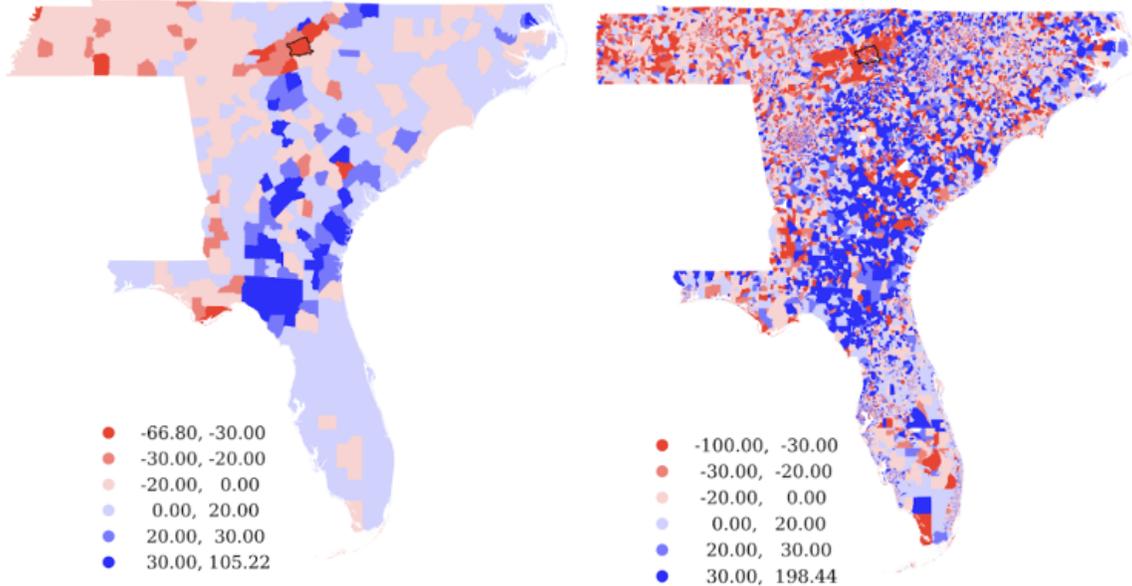

Figure 1: Mobility change (%) after versus before Hurricane Helene: Left – county level; Right – Census Block Group level.

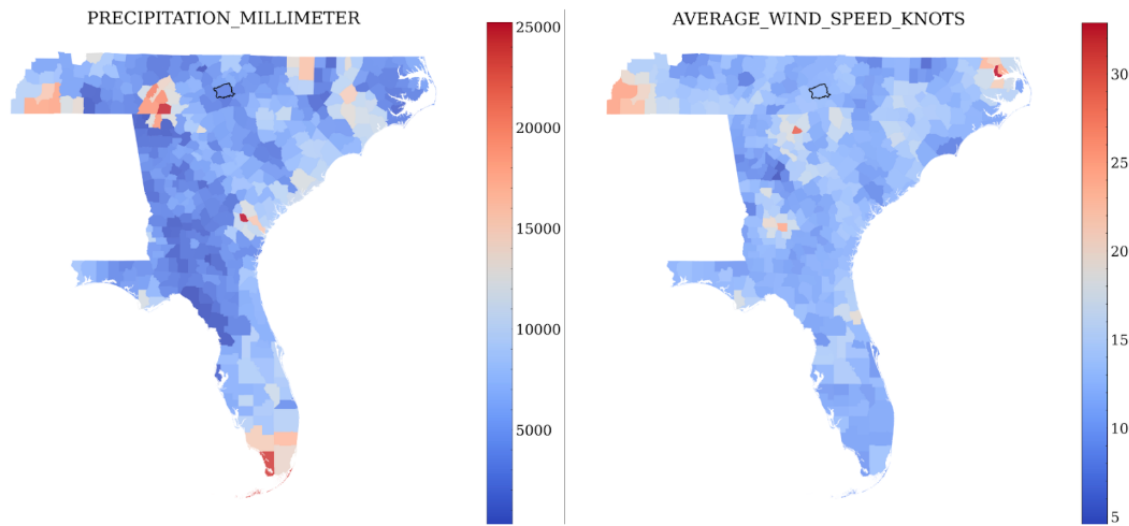

Figure 2: County-level weather conditions: average precipitation and average wind speed during Hurricane Helene.

Spatial visualizations reveal clear geographic variation in post-hurricane mobility. At a regional scale, much of the Southeast region experienced reductions in travel, with particularly sharp declines across large portions of Florida. Finer-resolution mapping reveals substantial heterogeneity within states: while many counties experienced mobility losses, adjacent areas sometimes showed sharp increases in mobility. This can suggest localized evacuation or early recovery activities. These spatial contrasts can also highlight potential differences in how local social and infrastructural conditions shape travel responses. We also see that weather exposure varied across the region. Precipitation was highest in several inland areas of Georgia and the Carolinas, while coastal zones experienced moderate rainfall. Stronger winds concentrated along parts of the Atlantic coast and the Carolinas, with generally weaker winds inland.

*Public Sentiment Analysis with Social Media Data*

Before conducting regression analysis, we also visualized the spatial distribution of average sentiment scores across the study area (Figure 3) based on Tweets we collected during the study period. Warmer (red) colors in the resulting maps represent more positive sentiment, while cooler (blue) colors represent more negative sentiment. `Overall, sentiment was skewed positively across most cities, with particularly strong concentrations of positive scores in parts of Florida. In contrast, many cities in Georgia exhibited clusters of negative sentiment, indicating a greater psychological strain or distress in those areas. Sentiment patterns were more mixed in other states, highlighting regional variation in public attitudes toward the hurricane. For instance, cities like Palatka, FL (average sentiment score = 1) and Thomasville, GA (0.667) showed more positive sentiment, indicating relative mental stability during the event. Conversely, places such as Douglas and Dublin, FL, recorded strongly negative sentiment scores (−1), suggesting that users there were more deeply affected by the hurricane's impacts.`

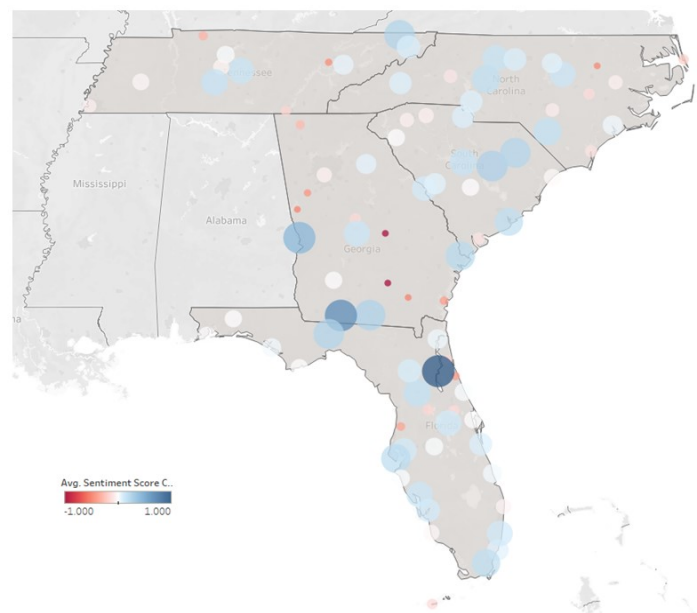

Figure 3: Average sentiment distribution over the study period

These visualizations offer an initial view of how post-hurricane mobility reflects both environmental exposure and social context. We examine the relationship between mobility disparity through contextualizing them in regression analysis while controlling for sociodemographic factors, storm conditions, evacuation orders, and public sentiments. Specifically, we ask two questions: (1) How do post-hurricane mobility patterns reflect community vulnerability and adaptive capacity? and (2) How do sociodemographic conditions and public sentiment factors shape the direction and extent of mobility change? To answer these questions, we set up two regression models to examine the factors that influence post-hurricane mobility. The first model uses robust linear regression to analyze factors associated with the *number of trip counts* within the two days following the hurricane's landfall. The second model uses ordered logistic regression to assess *the likelihood of increased mobility*. Both models control for related sociodemographic, economic, spatial, and weather-related predictors, as well as government-issued evacuation orders and county-level sentiment scores derived from Twitter data.

*Regression Model on Mobility Pattern after Hurricanes*

Regression results from the robust linear regression (Table 1) indicate that several social and spatial factors influence travel patterns following hurricanes. Higher population density was strongly associated with lower mobility. This pattern may reflect multiple dynamics. In dense urban areas, congestion and cascading failures in transit and road networks can exacerbate the impact of storm disruptions, significantly constraining post-disaster travel. At the same time, residents of denser neighborhoods can also possibly rely less on private vehicles and more on public transit or short local trips (Ewing & Cervero, 2010); when these alternative modes are disrupted, overall mobility can fall more steeply. At the same time, rural counties exhibited reduced mobility, which is consistent with their more limited transportation networks, longer travel distances, fewer evacuation routes, and lower adaptive capacity (Huang et al., 2016). These results highlight how both the vulnerabilities of dense urban networks and the scarcities of rural infrastructure restrict adaptive movement during disasters.

Regarding the sociodemographic conditions, we see that census block groups with lower median incomes were associated with reduced trips, which underscores how economic hardship can potentially constrain recovery-related travel. Census block groups with a higher percentage of Black population experienced significantly lower mobility. This can signal the presence of structural barriers to safe movement during the immediate aftermath of the disaster (Chakraborty et al., 2019; Elliott & Pais, 2006; Huang et al., 2016). We also see that communities with older populations (65 years and older) showed reduced mobility. This suggests that areas with a higher proportion of older adults made fewer trips during the study period. Several mechanisms can potentially explain this pattern. First, functional limitations and chronic conditions can reduce routine travel and increase trip costs and risks for older adults, especially when infrastructure or services are disrupted (Satariano et al., 2012). Second, after the hurricane landed, evacuation and disaster-related travel can pose elevated health risks for this group (e.g., ED visits, post-event institutionalization); therefore, older adults and care facilities sometimes opt for shelter-in-place when feasible, which reduces observed mobility (Hua et al., 2024). Third, older residents rely more on formal or informal assistance and specialized transportation, such as paratransit and medical transport, rather than private vehicle travel (Phraknoi et al., 2023). When these systems pause or scale back during an event, trips fall even if needs rise. In addition, preparedness gaps among older adults, lower rates of drills, kit readiness, and evacuation planning, can result in delayed or foregone movement under time pressure (Eisenman et al., 2007; Liao & Hu, 2025). A measurement caveat also leans in the same direction: smartphone-based mobility datasets underrepresent older adults and other disadvantaged groups, which can bias mobility downward in older communities even after controls (Z. Li et al., 2024).

Education levels and the share of the Hispanic/Latino population were both positively associated with higher mobility, possibly reflecting occupational travel or resource-seeking behavior. For example, communities with more college-educated residents may have greater access to resources, vehicles, and information networks, which allow them to have more trips even under disrupted conditions (Howe et al., 2015; Sadri et al., 2017). On the other hand, communities with a higher Hispanic/Latino population share are also linked to increased mobility. Unlike capacity-driven mobility, this seemingly similar high travel amount may reflect a different set of occupational demands, particularly a concentration in service and essential industries that require in-person work, as well as resource-seeking behavior during recovery (Zoraster, 2010; Esmalian et al., 2021).

Recall that we found a significant and negative association between the travel amount and the percentage of the Black population, which is the opposite of what we see with the percentage of the Hispanic/Latino population. We note that these differing directions of findings may suggest a greater need to contextualize mobility-related resilience or vulnerability (Deng et al., 2021). For some groups, economic necessity may drive greater mobility during disasters, while others encounter structural barriers that restrict movement and affect their ability to respond (B. Li et al., 2024).

Table 1: Coefficient of Robust Linear Regression Model on the Number of Trips during Hurricane Helene

| Avg_Trips during Helene | Coefficient | Robust std. err. | P>t | [95% conf. interval] | |
|---|---|---|---|---|---|
| **pop_density_scaled** | -0.0213874 | 0.0024751 | *** | -0.0262389 | -0.0165359 |
| **republican_r** | 0.0025603 | 0.0007291 | *** | 0.0011312 | 0.0039893 |
| **over_65_r** | -0.0026465 | 0.0006039 | *** | -0.0038301 | -0.0014629 |
| **rural_population_r** | -0.0010173 | 0.0001709 | *** | -0.0013524 | -0.0006823 |
| **hispanic_latino_r** | 0.0029518 | 0.0003888 | *** | 0.0021897 | 0.0037138 |
| **black_non_hispanic_r** | -0.0031762 | 0.0003076 | *** | -0.0037791 | -0.0025732 |
| **education_degree_r** | 0.004397 | 0.0006305 | *** | 0.0031611 | 0.0056328 |
| **median_income_scaled** | -0.0136801 | 0.0032471 | *** | -0.0200446 | -0.0073155 |
| **ln_prec** | 0.0415229 | 0.0063087 | *** | 0.0291573 | 0.0538885 |
| **ln_wind** | 0.1509751 | 0.0499113 | *** | 0.0531449 | 0.2488052 |
| **1. evacuation** | 0.057569 | 0.015139 | *** | 0.0278953 | 0.0872428 |
| sentiment_score_county_average | 0.0376824 | 0.0215626 | * | -0.0045821 | 0.0799469 |
| _cons | -0.7266395 | 0.0800076 | *** | -0.8834609 | -0.5698181 |

Note: Model Number of obs: 20,324;F(12, 20311) = 124.59;; Prob > F= 0; R-squared=0.0203; Root MSE= 1.1583 ; Statistical significance: *** indicates p < 0.01, ** indicates p < 0.05, and * indicates p < 0.10, based on two-tailed tests.

Consistent with our expectations, we find that weather conditions exert a strong association with post-hurricane travel behavior. Both heavier rainfall and higher wind speeds are positively associated with increased movement. This could reflect that residents were relocating for safety, seeking resources, or adjusting routines in response to hazardous conditions. County-level evacuation orders show a particularly strong positive association, which highlights their critical role in prompting large-scale mobility during disasters. While weaker in magnitude, the public sentiment score is also positively related to travel, suggesting that higher levels of optimism or community engagement at the county level may modestly encourage mobility, possibly through greater social coordination or perceived capacity to act.

From the findings above, we can see that the mere number of trips in post-hurricane travel cannot be interpreted as a uniform marker of resilience. In communities with lower income, rural, higher percentages of Black population, constrained mobility can often reflect structural barriers, such as limited transportation options, financial constraints, or restricted access to recovery resources, that hinder evacuation or relief-seeking. By contrast, reduced travel in other contexts may signal the capacity to remain safely sheltered, adequate preparation, or access to local resources. These divergent pathways together underscore the importance of situating mobility patterns within their social and geographic context when evaluating disaster resilience and vulnerability.

### *Regression Model on Increased Mobility: Coping Mechanism or Crisis Response?*

To further explore these dynamics, we next examine cases where mobility increased to examine whether such movement reflects adaptive coping strategies or acute responses to crisis. An ordered logistic regression assessed factors associated with increased travel post-hurricane (Table 2). Results show that evacuation orders had the strongest positive effect, meaning that communities that received evacuation orders are more likely to increase travel amount. This finding confirms the central role of government-issued emergency response guidance in shaping adaptive behavior in the aftermath of disasters. Population density and median income were also positively associated with a greater likelihood of mobility increases, indicating that residents in urban and wealthier areas may have greater flexibility and resources to travel. Higher educational attainment was associated with lower odds of increased travel. This pattern may reflect stronger risk awareness, allowing individuals to make informed decisions about staying put rather than relocating unnecessarily. It may also signal a greater capacity to shelter in place, including access to safe housing, resources, and reliable information, which reduces the need for post-hurricane movement. Wind intensity reduced the likelihood of increased travel, which suggests that damage or perceived danger discouraged movement. We again observe that the public sentiment score from Twitter data was positively associated with increased travel, suggesting that higher optimism or lower public distress was linked to greater mobility.

We found that the percentage of the Black population was associated with lower odds of increased mobility, reinforcing consistent racial disparities across models. As discussed above in the Model 1 results, these neighborhoods recorded fewer trips overall, suggesting that structural barriers, such as limited transportation access, financial constraints, or gaps in recovery resources, suppressed movement during the immediate response period. Building on this, our Model 2 shows that Black populations had lower odds of increasing mobility from previous travel amounts after the hurricane. This provides consistent evidence that Black-majority communities can experience constrained mobility in the aftermath of the hurricane, and points to a compounded disparity: Black communities not only began from a position of reduced travel, but were also less able to adapt their mobility in the face of escalating crisis demands.

To test whether income moderates the mobility gap associated with Black communities, we included an interaction term between the Black population share and income. From the result, we see that the positive coefficient indicates that higher income reduces the negative association between the Black share and increased mobility. This suggests that economic resources can partly offset mobility constraints in Black communities. When resources are available, as measured through an increase in income, Black communities are more equipped to adjust their travel. This finding reinforces that the reduced mobility observed in low-income Black areas reflects structural barriers rather than a greater capacity to shelter in place.

Table 2: Coefficient of the Ordered Logistic model estimating the odds ratio of increased travel during Hurricane Helene versus before Hurricane Helene

| y= binary_change_mobility | Coefficient | Robust std. Err. | P>t | [95% conf. interval] | |
|---|---|---|---|---|---|
| **pop_density_scaled** | 0.0231876 | 0.0042421 | *** | 0.0148733 | 0.0315019 |
| republican_r | -0.0002492 | 0.0013782 | 0.857 | -2.95E-03 | 0.0024521 |
| over_65_r | 0.0016439 | 0.0011678 | 0.159 | -0.000645 | 0.0039328 |
| rural_population_r | -0.000487 | 0.0005127 | 0.342 | -0.0014919 | 0.000518 |
| hispanic_latino_r | -0.0021159 | 0.0016492 | 0.2 | -0.0053483 | 0.0011166 |
| **Black** | -0.0084586 | 0.0013106 | *** | -0.0110272 | -0.0058899 |
| **income_black** | 0.0006099 | 0.0002511 | ** | 0.0001178 | 0.001102 |
| income_hispanic | 0.0001801 | 0.0002249 | 0.423 | -0.0002606 | 0.0006209 |
| **education_degree_r** | -0.0107956 | 0.0011619 | *** | -0.0130729 | -0.0085182 |
| **median_income_scaled** | 0.0259356 | 0.0077975 | *** | 0.0106526 | 0.0412185 |
| ln_prec | -0.0053458 | 0.0236213 | 0.821 | -0.0516426 | 0.040951 |
| **ln_wind** | -0.4035419 | 0.0858707 | *** | -0.5718454 | -0.2352384 |
| **1.evacuation** | 0.3232111 | 0.0445125 | *** | 0.2359682 | 0.4104539 |
| sentiment_score_county_average | 0.1795518 | 0.0877324 | ** | 0.0075994 | 0.3515042 |
| /cut1 | -1.407666 | 0.2859966 | | -1.968209 | -0.8471228 |

Number of obs = 20,324; Wald chi2(14) = 288.45; Prob > chi2 = 0.0000
R-squared = 0.0109; Log pseudolikelihood = -13825.443;
Statistical significance: *** indicates p < 0.01, ** indicates p < 0.05, and * indicates p < 0.10, based on two-tailed tests.

The findings from above demonstrate that increased post-hurricane mobility does not uniformly indicate adaptive resilience; it may instead reflect evacuation under duress or urgent efforts to meet recovery needs. Communities with high travel volumes, particularly when driven by evacuation orders or severe weather, often respond to acute threats and instability rather than exercising choice. Conversely, limited increases in movement can signal two very different realities: structural barriers that constrain mobility in resource-poor or marginalized communities, or sufficient preparedness and shelter-in-place capacity in households with greater resources and safer housing.

We recognize that distinguishing between these pathways is crucial for interpreting mobility as either a marker of vulnerability or resilience, as well as the challenges associated with doing so. While individual-level survey data can provide valuable insights into the motivations and constraints behind mobility decisions, collecting such information is often challenging in post-disaster contexts, where trauma and the immediate toll of recovery already weigh heavily on affected residents after disaster events.

**Discussion and Conclusions**

Post-hurricane mobility patterns consistently reveal disparities across race, income, and other social dimensions, raising fundamental equity concerns in evacuation and recovery processes (Bethel et al., 2011; Deng et al., 2021; X. Li et al., 2024; T. Logan et al., 2019; Swanson & Guikema, 2024). Understanding the context of mobility needs and constraints after a hurricane is crucial for understanding who can move, who remains in place, and why. These distinctions carry direct implications for resilience, vulnerability, and equity in disaster response. Through evidence from Hurricane Helene, our findings demonstrate that post-hurricane mobility shifts cannot be interpreted in isolation from their social and environmental context. Using regression models that integrate GPS human mobility data, social media sentiment, and hurricane severity metrics, we find that mobility patterns following Hurricane Helene were shaped not only by storm conditions but also by structural inequalities. Communities with lower incomes, rural areas, and a higher percentage of Black populations exhibited the steepest declines in mobility, likely reflecting resource constraints and infrastructure barriers, while wealthier, urban, and higher-education areas sustained greater flexibility. Public sentiment was positively associated with increased mobility, which indicates that higher optimism or lower distress at the county level corresponded with greater travel. This relationship suggests that collective outlook and social engagement may play a role in shaping movement during the immediate aftermath of the hurricane.

We argue that reduced mobility did not always indicate vulnerability; some communities likely remained stable due to preparedness and the capacity to shelter in place, while increased trips sometimes signaled both evacuation-driven displacement and continued access to functional networks. Our findings also suggest that communities with a larger share of older adults exhibit reduced mobility. While this likely reflects real limitations and dependence on external support systems, it is important to acknowledge potential data limitations. Smartphone-based mobility datasets can underrepresent older populations and other disadvantaged groups, which may bias measured mobility downward in these communities, even after controlling for other factors. This reinforces the need for caution when interpreting digital mobility data, as patterns attributed to vulnerability and dependence may partly reflect gaps in representation. Our findings also call attention to the need for more inclusive data systems that capture the mobility realities of groups who are often underrepresented in digital records, such as older adults and individuals with limited access to technology.

Mobility equity should be a central part of resilience strategies for urban planners, transportation planners, and emergency response agencies. Evacuation routes, transit services, and recovery resources must be designed to reach communities that often face the greatest barriers, including those with lower incomes, older adults, residents of rural areas, and historically marginalized communities with higher percentages of Black populations. Targeted steps such as pre-event enrollment in assisted transportation, continuity plans for caregivers and essential medications, and clear risk communication tailored to people with fewer options for movement can help reduce these risks. Our findings also highlight the importance of public sentiment; places where people expressed more optimism or community engagement exhibited greater mobility after the storm, suggesting that social ties and trust can influence how people respond. Addressing both the structural barriers in post-disaster mobility and the social conditions that influence risk-adaptive actions is key to building more just and effective disaster recovery planning.

**Conflict of Interest:** The authors declare that they have no competing interests related to this study.